\documentclass[preprint,12pt]{elsarticle}
\usepackage{graphics}

\usepackage{amssymb}

\journal{Physics Letters A}

\usepackage[english]{babel}
\renewcommand{\Re}{\mathop{\mathrm{Re}}\nolimits}

\begin{document}

\begin{frontmatter}



\title{On the motion of high energy wave packets and the transition radiation by ``half-bare'' electron}


\author[sh1]{N.F.~Shul'ga\corref{cor1}}
\ead{shulga@kipt.kharkov.ua} \cortext[cor1]{Corresponding author}

\author[syshch]{V.V.~Syshchenko}
\ead{syshch@bsu.edu.ru}

\author[sh1]{S.N.~Shul'ga}

\address[sh1]{Akhiezer Institute for Theoretical Physics,
National Science Center ``Kharkov Institute of Physics and
Technology'', Akademicheskaya st., 1, Kharkov 61108, Ukraine}
\address[syshch]{Belgorod State University, Pobedy st., 85, Belgorod 308015, Russian Federation}

\begin{abstract}
The problem of the motion of high-energy wave packets combined of
free electromagnetic waves is considered. It is demonstrated that
the transformation of such packets to the packet of spherically
diverging waves happens on long distances along the packet's
motion direction, that substantially exceed the radiated
wavelength. The transition radiation by the ``half-bare''
ultrarelativistic electron is considered. It is demonstrated that
the transition radiation by such an electron on the targets
located inside and outside the coherence length of the radiation
process would be substantially different.
\end{abstract}


\end{frontmatter}


\section{Introduction}
\label{}

Moving electron is the charge and the eigenfield (Coulomb
field) moving together with it. Changing the electron's trajectory
disturbs that field. The disturbance of the field could be treated
as a packet of free plane electromagnetic waves. On large
distances from the region where the acceleration had happened the
packet transforms to the packet of diverging waves (the radiation
field). For non-relativistic particles that happens on the
distances of order of the length $\lambda$ of radiated wave
\cite{LL2}. High energies make the stabilizing influence to wave
packets that leads to a substantial increase of the length on which the
packet's transformation takes place. This length could have
macroscopic size, exceeding not only interatomic distances in
matter, but also the size of the target and just the size of the
experimental installation (detector). Hence it is important to
know the behavior of such high-energy packets of electromagnetic
waves in the region where that transformation happens. The present
article is devoted to the examination of this problem.

Primarily, we consider the motion of Gaussian packet combined of
plane waves with directions of the wave vector $\mathbf k$ close
to each other. It is demonstrated that the shape of such a packet
changes on the lengths that substantially exceed the wavelength
$\lambda = 1/|\mathbf k|$ corresponding to the given absolute
value of the wave vector $|\mathbf k|$.

Then we consider the motion of the wave packet that coincides in
the time moment $t=0$ with the eigenfield of the ultrarelativistic
electron. It is demonstrated that the last packet also conserves
its shape for a long time interval. Fourier component of this
packet with the wavelength $\lambda$ changes only on the distances
$z$ along the packet's direction of motion that exceed the length
$2\gamma^2\lambda$, where $\gamma$ is the electron's Lorentz
factor. This length coincides with the coherence length of the
radiation process of the relativistic electron $l =
2\gamma^2\lambda$ \cite{TerMik, AhSh}.

The problem of special interest is the radiation under sharp (at
the time moment $t=0$) changing of the ultrarelativistic
electron's velocity [3~-~5]. We demonstrate
that the packets of electromagnetic waves arising in this case are
close in their structure to the packets considered above. However,
their manifestations in the direction of the initial and final
motion of the electron are substantially different. Namely, on the
distances $z < 2\gamma^2\lambda$, Fourier components with the
wavelength $\lambda$ of the packet moving along the direction of
the initial electron motion will practically coincide with the
Fourier components of the initial packet and, consequently, to the
Fourier components of the Coulomb field of the electron moving in
the initial direction without scattering. Oppositely, in the final
electron's direction of motion, the field of the packet of free
waves will screen the particle's eigenfield. The electron under
such conditions was called in \cite{Feinberg} as ``half-bare
particle'', that is the particle whose specific Fourier components
of the surrounding field are practically absent for a long time.
We put attention to that the transition radiation by such
particles and wave packets on the targets placed on the distances
from the point of scattering larger and smaller than
$2\gamma^2\lambda$ would be substantially different. The
corresponding experiment would permit to observe direct
manifestation of the ``half-bare'' electron and the process of its
dressing.

Let us note that for the charged particle the Gauss theorem is
applicable, according to which the number of force lines of the
electromagnetic field surrounding the electron does not change
with time \cite{LL2}. Under this the radiation process by electron
can be presented as bending of these force lines [6~-~10]. Such a
concept of radiation process relates to the complete
electromagnetic field surrounding the electron. However, it does
not contain such characteristics of the radiation process as
coherence length and wave zone which are connected with determined
Fourier components of this field. The term ``half-bare electron''
relates also to a determined Fourier component of the field
surrounding the electron which is defined by the wavelength
$\lambda$. So, the analysis of a space-time evolution of these
Fourier components (wave packets) gives us a supplement for the
picture of evolution of complete field surrounding the electron
which is in accelerated motion.

We use the system of units in which the speed of light in vacuum
is taken equal to the unit: $c=1$.

\section{Gaussian packet}
\label{} The scalar potential of the packet of free
electromagnetic waves could be expressed in the form of the
following Fourier decomposition:
\begin{equation}
\varphi (\mathbf r, t) = \int \frac{d^3q}{(2\pi)^3} e^{i(\mathbf q
\mathbf r - qt)} C_q, \label{1}
\end{equation}
where $C_q$ are the coefficients of the decomposition, $q =
|\mathbf q|$. Consider at first the behavior of the packet
combined at $t=0$ of plane waves with the wave vectors $\mathbf k$
directed closely to some given direction (the $z$-axis). Supposing
for simplicity that the distribution of the waves over directions
of the vector $\mathbf k$ is Gaussian at $t=0$, let us write the
potential (\ref{1}) in the form
\begin{equation}
\varphi_k (\mathbf r, 0) = \frac{1}{\pi \Delta^2} \int d^2
\vartheta e^{-\vartheta^2/\Delta^2} e^{i\mathbf k \mathbf r} ,
\label{2}
\end{equation}
where $\vartheta$ is the angle between $\mathbf k$ and the
$z$-axis, and $\Delta^2$ is the mean square value of the angle
$\vartheta$, $\Delta^2\ll 1$. Coefficients $C_{\mathbf q}$ of a
such packet have the form
\begin{equation}
C_{\mathbf q} = (2\pi)^3 \int \frac{d^2\vartheta}{\pi\Delta^2}
e^{-\vartheta^2/\Delta^2} \delta(\mathbf k - \mathbf q), \label{3}
\end{equation}
where $\delta(\mathbf k - \mathbf q)$ is the delta-function. In
this case, according to (\ref{1}),
\begin{equation}
\varphi_k (\mathbf r, t) = \frac{1}{1+ikz\Delta^2/2} \exp\left\{
ik(z-t) - \frac{(k\rho\Delta/2)^2}{1+i(kz\Delta^2/2)} \right\},
\label{4}
\end{equation}
where $\rho$ is the transverse (in relation to the $z$-axis)
component of $\mathbf r$.

Eq. (\ref{4}) demonstrates that under $kz\Delta^2/2 \ll 1$
\begin{equation}
\varphi_k (\mathbf r, t) \approx \exp\left\{ ik(z-t) -
(k\rho\Delta/2)^2 \right\}, \label{5}
\end{equation}
and under the condition $kz\Delta^2/2 \gg 1$
\begin{equation}
\varphi_k (\mathbf r, t) \approx - \frac{2i}{kz\Delta^2}
\exp\left\{ ik(z-t) + ik \frac{\rho^2}{2z} -
\frac{\rho^2}{z^2\Delta^2} \right\}. \label{6}
\end{equation}
In the case $z\gg\rho$ the last formula could be written in the
form of diverging wave:
\begin{equation}
\varphi_k (\mathbf r, t) \approx - \frac{2i}{kr\Delta^2}
\exp\left\{ ik(r-t) - \frac{\rho^2}{z^2\Delta^2} \right\},
\label{7}
\end{equation}
where $r=\sqrt{\rho^2+z^2}\approx z + \rho^2/2z$.

So, on the distances $z$ from the center of the packet that
satisfy the condition
\begin{equation}
kz\Delta^2/2\ll 1, \label{8}
\end{equation}
the shape of the packet (\ref{4}) coincides with the packet's
shape at $t=0$. Only on the distances $z$ that satisfy the
condition
\begin{equation}
kz\Delta^2/2 >1, \label{9}
\end{equation}
the transformation of the packet of plane waves (\ref{4}) into the
packet of diverging spherical waves happens.

In the theory of radiation of electromagnetic waves, the spatial
region where the field of moving charges acquires the form of
spherically diverging waves, is called as {\it wave zone} (see,
e.g. \cite{LL2, Jackson}). Particularly, for non-relativistic
charged particles the wave zone begins just on the distances from
the radiating system that exceed the radiated wavelength (see
\cite{LL2}). Condition (\ref{9}) demonstrates, however, that
under $\Delta^2\ll1$ the formation of the wave zone takes
place not on the
distances $z>\lambda$, like in the problem of radiation of the
non-relativistic particle, but on the distances
\begin{equation}
z > 2\lambda /\Delta^2, \label{10}
\end{equation}
which are much larger than the wavelengths $\lambda = 1/k$, of which the
packet is composed (\ref{4}). For small values of $\Delta^2$
the length $z = 2\lambda /\Delta^2$ could reach macroscopic sizes.

\section{Approximation of Coulomb field by the packet of plane waves}
\label{}

Such problem arises in the equivalent photons method (or the
method of virtual quanta) when the Coulomb field of relativistic
electron is replaced at some specific time moment ($t=0$) by the
packet of free electromagnetic waves. Indeed, the Fourier
decomposition of the electron's Coulomb field could be written in
the form
\begin{equation}
\varphi_c (\mathbf r, t) = \Re \int \frac{d^3k}{(2\pi)^3}
e^{i(\mathbf k \mathbf r - \mathbf k \mathbf v t)} C_k^c,
\label{11}
\end{equation}
where $\mathbf v$ is the electron's velocity directed along the
$z$-axis, and
\begin{equation}
C_k^c = \frac{8\pi e \Theta (k_z)}{k_\perp^2 + k_z^2 /\gamma^2}.
\label{12}
\end{equation}
Here $\gamma$ is the electron's Lorentz factor, $k_z$ and $\mathbf
k_\perp$ are the components of the vector $\mathbf k$, parallel
and orthogonal to the $z$-axis, $\Theta (k_z)$ is the Heaviside's
step function.

It is supposed in the equivalent photons method that at $t=0$ the
packet (\ref{1}) composed of free electromagnetic waves coincides
with the electron's Coulomb field moving with the velocity
$\mathbf v$ [11~-~13]. That corresponds
to Fourier decomposition (\ref{1}) with the coefficients $C_q =
C_k^c$.

For $\gamma\gg 1$ the main contribution to (\ref{1}) would be made
by the values $\mathbf q = \mathbf k$ which directions are close
to the direction of the electron's velocity $\mathbf v$. Taking
this into account, the packet (\ref{1}) could be written in the
form
\begin{equation}
\varphi (\mathbf r, t) = \Re \int_0^\infty dk\, \varphi_k(\mathbf
r, t) , \label{13}
\end{equation}
where
\begin{equation}
\varphi_k (\mathbf r, t) = \frac{2}{\pi} \exp\left[ ik(z-t)\right]
 \int_0^\infty \frac{\vartheta
d\vartheta}{\vartheta^2 + \gamma^{-2}} J_0(k\rho\vartheta)\, e^{
-ikz\,\vartheta^2/2}  . \label{14}
\end{equation}
Here $\vartheta$ is the angle between $\mathbf k$ and $\mathbf v$
($\vartheta\ll 1$), and $J_0(x)$ is the Bessel function.

The function $\varphi_k (\mathbf r, t)$ has the same structure as
the function (\ref{4}) corresponding to Gaussian distribution of
the vectors $\mathbf k$ over the angles $\vartheta$. Namely, if
$kz\vartheta^2/2\ll 1$, the main contribution to the integral
(\ref{14}) is made by the values $\vartheta\sim\gamma^{-1}$ and
\begin{equation}
\varphi_k (\mathbf r, t) \approx \frac{2}{\pi} K_0(k\rho/\gamma)
\, e^{ik(z-t)} , \label{15}
\end{equation}
where $K_0(x)$ is the modified Hankel function.
In this case after integration over $k$ in (\ref{13}) we find that
\begin{equation}
\varphi (\mathbf r, t) \approx \frac{e}{\sqrt{(z-t)^2 +
\rho^2/\gamma^2}} . \label{16}
\end{equation}
The main contribution to (\ref{13}) is made by the values $k\sim
\gamma/\rho$, hence Eq. (\ref{16}) is valid in the range of
coordinates $\rho$ and $z$ that satisfy the condition $z <
\gamma\rho$. In this range of coordinates the packet under
consideration moves with the velocity of light in the $z$-axis
direction.

So, on the distances $z\lesssim 2\gamma^2\lambda$ the considered
wave packet practically coincides with the initial one (at $t=0$).
Substantial transformation of the packet would happen only on the
distances
\begin{equation}
z > 2\gamma^2\lambda . \label{17}
\end{equation}
In this case for the evaluation of the integral in (\ref{14}) over
$\vartheta$ one could apply the method of stationary phase. As a
result of using of this method we find that
\begin{equation}
\varphi_k (\mathbf r, t) = -\frac{2i}{\pi} \frac{1}{ \vartheta_0^2
+ \gamma^{-2}} \frac{1}{kr} e^{ik(r-t)} , \label{18}
\end{equation}
where $r\approx z + \rho^2/2z$ and $\vartheta_0 = \rho /z$ is the
point of stationary phase of the integral (\ref{14}). We see that
the components (\ref{18}) of our packet have in the case under
consideration the form of diverging spherical waves. Under this condition the angle $\vartheta_0$ corresponds to the direction of radiation,
and the function before the diverging wave describes the angular
distribution of the radiation. So, the condition (\ref{17}) draws
out the wave zone in application to given problem.

The value $2\gamma^2\lambda$ presenting in the condition (\ref{17})
is known in the theory of radiation by ultrarelativistic particles
as the {\it formation length} or the {\it coherence length}
\cite{TerMik, AhSh}.

\section{Transition radiation by a ``half-bare electron''}
\label{}

High-energy packets of electromagnetic waves considered above
manifest themselves in many problems connected with bremsstrahlung
and diffraction radiation (see, e.g., \cite{ShFomin, ShDobr,
Naum}). Let us pay attention to some manifestations of such
packets in the problem of transition radiation arising after sharp
scattering of the high-energy electron on large angle.

The retarded solution for the potential of the electromagnetic
field after the scattering of the electron at the time moment
$t=0$ on large angle could be expressed in the following form
\cite{AhSh}:
\begin{equation}
\varphi (\mathbf r, t) = \Theta (r-t) \varphi_{\mathbf v} (\mathbf
r, t) + \Theta (t-r) \varphi_{\mathbf v'} (\mathbf r, t),
\label{19}
\end{equation}
where $\varphi_{\mathbf v} (\mathbf r, t)$ and $\varphi_{\mathbf
v'} (\mathbf r, t)$ are potentials of the Coulomb field of the
electrons moving all the time with the velocity $\mathbf v$ along
the $z$-axis and with the velocity $\mathbf v'$ along the
$z'$-axis, respectively. Eq. (\ref{19}) demonstrates that after
scattering of the electron at $t=0$ its eigenfield strips out and
after that transforms into the radiation field. In the direction
of the final particle's motion the electron's eigenfield arises
only in the region $r<t$ which is achieved by the signal about the
scattering act at $t=0$ (see Fig. 1, where the isolines of the
scalar potential (\ref{19}) are presented).

Consider the Fourier decomposition of (\ref{19}):
\begin{eqnarray}
\varphi (\mathbf r, t) = {e\over 2\pi^2} \Re \int {d^3 k\over k}
e^{i\mathbf k \mathbf r} \left\{ {1\over k-\mathbf k\mathbf v}
e^{-ikt} \right. \nonumber \\ + \left. {1\over k-\mathbf k\mathbf
v'} \left[ 1-e^{-i(k-\mathbf k\mathbf v')t} \right] e^{-i\mathbf
k\mathbf v' t} \right\} . \label{20}
\end{eqnarray}

The first term in this formula has the form of the packet of free
waves moving along initial direction of the electron's velocity $\mathbf v$. This packet coincides with the electron's eigenfield at $t=0$.
According to (\ref{17}), (\ref{18}), the transformation of the Fourier components of this packet with the wavelength $\lambda$ to the packet of diverging waves would happen on the distances $z > 2\gamma^2\lambda$. On smaller distances the packet of waves with the given value of $|\mathbf k|$ would be close to the initial one.

The length $l = 2\gamma^2\lambda$ on which the formation of the
wave zone takes place could have macroscopic size. For example,
for the electrons of energy 50 MeV in the range of wavelengths
$\lambda\sim 10^{-1}$ cm this length is about 20 m (the measuring
technique in such conditions is developed today --- see, e.g.
\cite{Naum, Shibata} ). So in the frames of that length one could
arrange a thin target (see the target in Fig.1 which is arranged
along the z-axis at $z < 2\gamma^2\lambda$) and examine the
``transition radiation'' of the considered packet (reflection of
the waves, their passage through target etc.). The characteristics
of such ``transition radiation'' practically would not differ from
the characteristics of the transition radiation of the electron
moving in the same direction (however, the electron in the packet
under consideration is absent). But if the target would be located
on the distance $z
> 2\gamma^2\lambda$ (see dashed-line box in Fig.1), the features of the considered ``transition
radiation'' would change due to the changing of the packet's shape
(formation of the diverging waves).

The second term in (\ref{20}) describes the field surrounding the
electron after its scattering at $t=0$, when its velocity became
equal to $\mathbf v'$. This field consists of the electron's
eigenfield moving with the velocity $\mathbf v'$ (the first term in
square brackets in (\ref{20})) and the packet of free waves moving
in the direction of $\mathbf v'$ coinciding at $t=0$ with the
opposite sign with Coulomb field of the electron (the second term
in square brackets).

As it was demonstrated above, transformation of the packet of
plane waves to the packet of diverging waves takes place on
the distances $z' \sim 2\gamma^2 \lambda$, where the axis $z'$ is
directed along $\mathbf v'$. During the time interval $t$ over
which the electron passes that distance, the substantial
cancellation of the terms in the square brackets in (\ref{20}) takes place.
This mean that the electron stays on that distance in a
``half-bare'' state: the Fourier components with the wave vector
$\mathbf k$ of its surrounding field would be suppressed comparing
to the case $z' > 2\gamma^2 \lambda$. Transition radiation of the electron with such field
(``half-bare'' electron) on the target located on the distance $z'
< 2\gamma^2 \lambda$ from the point of scattering (see the target on Fig.1 which is arranged along the $z'$-axis) would be
suppressed in comparison to the case $z' > 2\gamma^2 \lambda$.

\begin{figure}
\begin{center}
\includegraphics[scale=0.7]{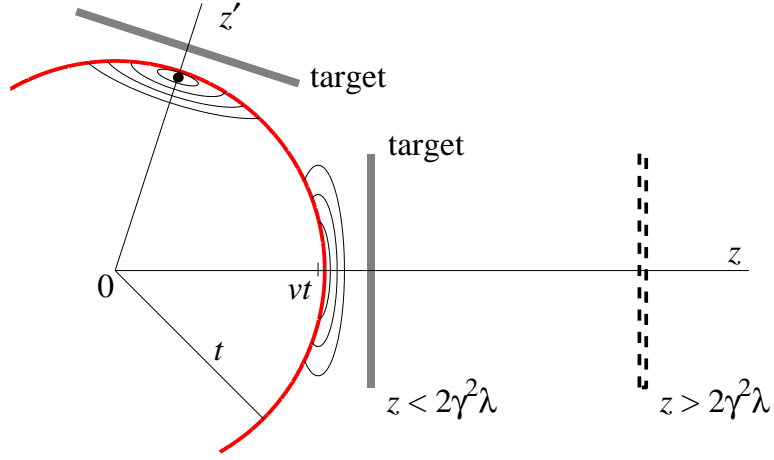}
\end{center}
\caption{Equipotential surfaces of (\ref{19}) and possible
positions of targets for producing of the transition radiation.}
\end{figure}

The results obtained are correct for sharp scattering of an electron at a large angle. Sharp scattering means that it takes place on the length which is much smaller than the coherent radiation length. At macroscopical values of the coherent length $l=2\gamma^2\lambda$ it can occur not only under scattering of an electron by an atom but also under its scattering by a magnet. The only condition required for this is that the size of a scatterer was small as compared with the coherent radiation length.

Note that bremsstrahlung arising under collisions of the
``half-bare'' electron with the atoms of the medium located in the
frames of the radiation formation length is suppressed comparing
to the case when the collisions happen out of that length
\cite{ShFomin, AhSh2, Feinberg2}. That lead, particularly, to such
effects as Landau-Pomeranchuk-Migdal effect of suppression of the
radiation by ultrarelativistic electrons in amorphous medium, the
effect of suppression of the coherent bremsstrahlung in crystals
and the effect of suppression of the radiation in thin layers of
substance (see recent reviews and monographs [19~-~22] devoted to
this topic, and the references therein). Experimental studies of
these effects were carried out during last years and are made at
present time on the accelerators of ultra high energies (see,
e.g., [21~-~23]). Examination of the process of transition
radiation by ``half-bare'' electron creates one more opportunity
for study of manifestations of such an electron under its
interaction with matter.

\section*{Acknowledgements}

This work is supported in part by the internal grant of Belgorod
State University.

\end{document}